\documentclass[twocolumn,english]{revtex4-1}
\usepackage[T1]{fontenc}
\usepackage[latin9]{inputenc}
\usepackage{amsmath}
\usepackage{graphicx}

\makeatletter
\usepackage{xcolor}
\usepackage[colorlinks,citecolor=red,linkcolor=blue]{hyperref}

\makeatother

\usepackage{babel}
\begin{document}
\title{Simple realization of the polytropic process with a finite-sized reservoir}
\author{Yu-Han Ma}
\email{yhma@gscaep.ac.cn}

\address{Graduate School of China Academy of Engineering Physics, No. 10 Xibeiwang
East Road, Haidian District, Beijing, 100193, China}
\begin{abstract}
In many textbooks of thermodynamics, the polytropic process is usually
introduced by defining its process equation rather than analyzing
its actual origin. We realize a polytropic process of an ideal gas
system when it is thermally contact with a reservoir whose heat capacity
is a constant. This model can deepen students' understanding of typical
thermodynamic processes, such as isothermal and adiabatic processes,
in the teaching of thermodynamics. Moreover, it can inspire students
to explore some interesting phenomena caused by the finiteness of
the reservoir. The experimental implementation of the proposed model
with realistic parameters is also discussed.
\end{abstract}
\maketitle
In conventional textbook of thermodynamics \citep{callen1960thermodynamics,cengel2011thermodynamics},
the polytropic process of the ideal gas is usually introduced as an
direct generalization of the isothermal and adiabatic processes. From
a teaching point of view, after studying the isothermal process equation
$PV=\mathrm{const}$ and the adiabatic process equation $PV^{\gamma}=\mathrm{const}$,
it seems easy for students to accept that there exists more practical
processes with $PV^{\xi}=\mathrm{const}$. Here, $P$ and $V$ are
the gas pressure and gas volume, respectively, $\gamma$ is the heat
capacity ratio, and $\xi$ is called the polytropic exponent . 

However, apart from the equation $PV^{\xi}=\mathrm{const}$ used to
define the polytropic process, the students are provided with little
or even no understanding of the specific characteristics, especially
the actual origin of such a process. It is well known that an isothermal
process can be achieved by quasi-statically expanding or compressing
the ideal gas when it is contacting with a heat reservoir with constant
temperature. In addition, when we quasi-statically compress or expand
the ideal gas which is thermally isolated, the adiabatic process is
achieved. It is therefore natural for one to ask what kind of realistic
thermodynamic process can be described by the polytropic process equation?
For a long time, the pursuit of such a question is hindered by the
direct generalization from the specific cases (isothermal and adiabatic
processes) to the polytropic process equation.

In this paper, we propose a simple model to realize the polytropic
process of the ideal gas. Following from the first law of thermodynamics
of energy conservation, the polytropic process equation is directly
obtained. We also discuss the finite size effect of the thermal reservoir
implied from this model.

\begin{figure}
\centering{}\includegraphics[width=8.5cm]{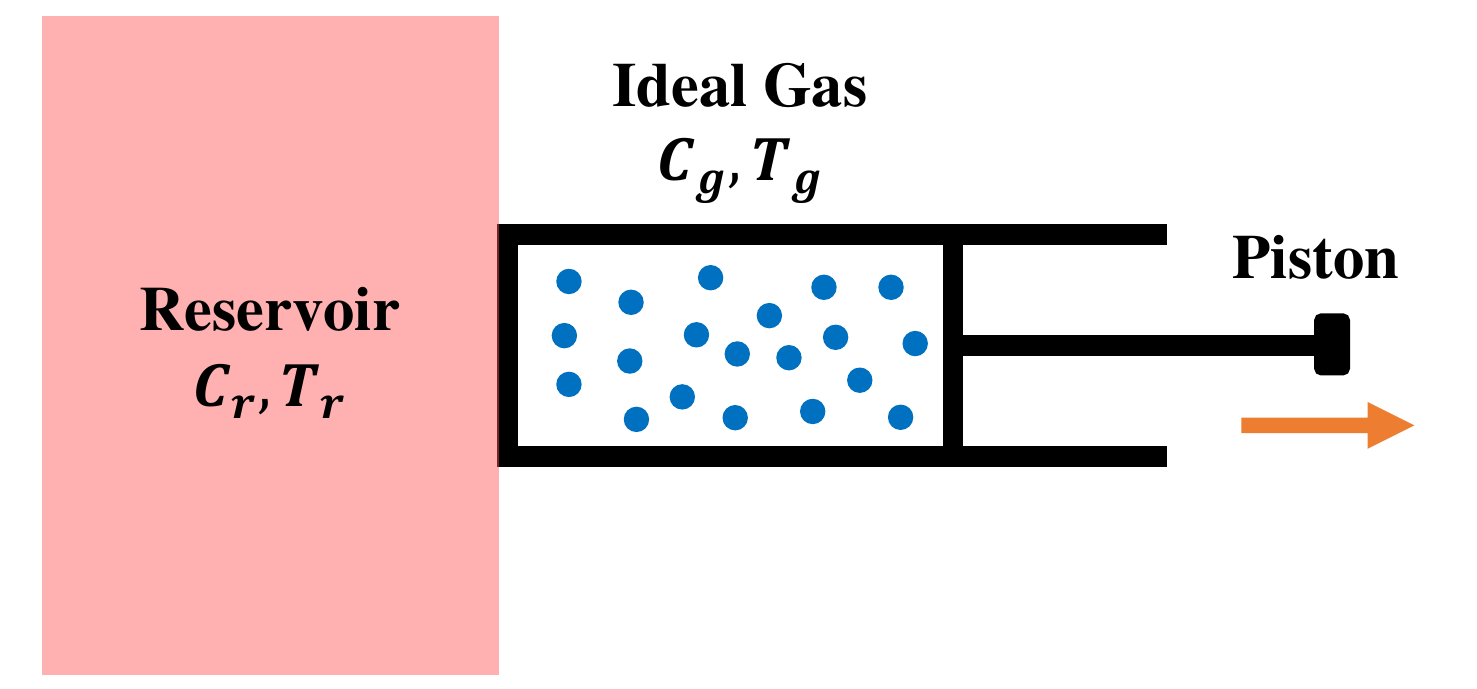}\caption{\label{fig:Schematic-diagram-of}Schematic diagram of the gas-reservoir
system. $T_{r}$ ($T_{g}$) and $C_{r}$ ($C_{g}$) are the temperature
and heat capacity at constant volume of the reservoir (gas), respectively.}
\end{figure}

\textit{Quasi-static energy transfer process in the gas-reservoir
system.--} As illustrated in Fig. \ref{fig:Schematic-diagram-of},
the whole system consists of an ideal gas system with temperature
$T_{g}$ and a thermal reservoir with temperature $T_{r}$. The gas
can be compressed or expanded through the piston. The thermal conductivity
between the gas and the reservoir is very well, which ensures the
gas is in thermal equilibrium with the reservoir all the times. In
this case, we have $T_{g}=T_{r}=T$. When the gas is quasi-statically
driven, the energy conservation of the gas in an infinitesimal process
follows as

\begin{equation}
dU_{g}=-PdV+\delta Q_{g},\label{eq:1st}
\end{equation}
where $U_{g}$ is the internal energy of the gas, $Q_{g}$ is the
heat absorbed by the gas from the reservoir. On the other hand, the
change of internal energy $U_{r}$ of the reservoir is

\begin{equation}
dU_{r}=\delta Q_{r}=C_{r}dT.
\end{equation}
Here, $C_{r}$ is the heat capacity at constant volume of the reservoir
and we have assumed that no work is applied to the reservoir. In the
following, we refer to the heat capacity at constant volume as the
heat capacity.

\begin{figure}

\centering{}\includegraphics[width=8.5cm]{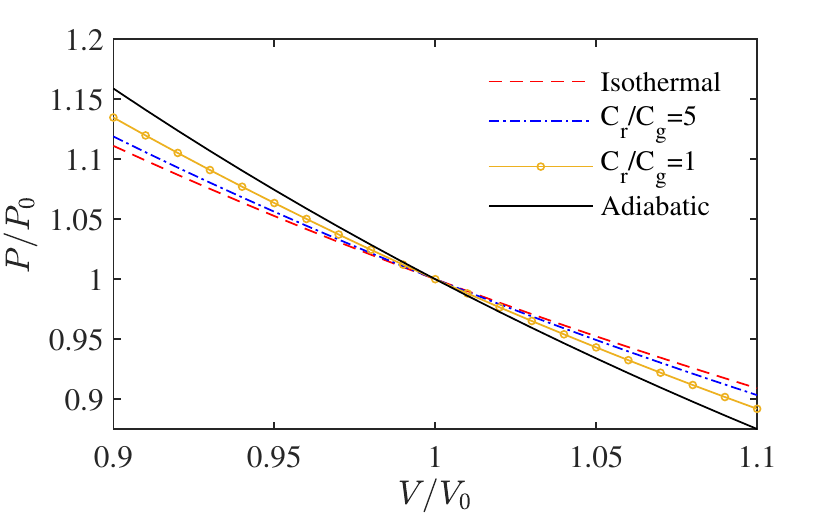}\caption{\label{fig:-diagram-of}$P-V$ diagram of the gas with different $C_{r}/C_{g}$.
The blue dash-dotted curve and yellow circled curve are plotted with
$C_{r}/C_{g}=5$ and $C_{r}/C_{g}=1$, respectively. The red dashed
curve represents the isothermal thermal process ($C_{r}\rightarrow\infty$)
while the black solid curve represents the adiabatic process ($C_{r}\rightarrow0$).
$P_{0}$ ($V_{0}$) is the initial gas pressure (volume) for a given
process. In this example, we use $\gamma=1.4$.}
\end{figure}

Noticing that the internal energy change of the ideal gas $dU_{g}=C_{g}dT$
with the heat capacity of the gas $C_{g}$, and $\delta Q_{g}=-\delta Q_{r}$,
we can rewrite Eq. (\ref{eq:1st}) as

\begin{equation}
\delta Q_{g}=C_{g}dT+PdV=-C_{r}dT.\label{eq:dQg}
\end{equation}
By substituting the ideal gas equation into Eq. (\ref{eq:dQg}), we
obtain the following differential equation

\begin{equation}
\left(C_{g}+C_{r}\right)dT=-PdV=-\frac{Nk_{B}T}{V}dV,\label{eq:DE}
\end{equation}
with the number of gas particle $N$ and the Boltzmann constant $k_{B}$.
With the assumption that the reservoir has constant heat capacity,
the solution of Eq. (\ref{eq:DE}) is straightforward obtained as

\begin{equation}
TV^{\left(\frac{\gamma-1}{1+C_{r}/C_{g}}\right)}=\mathrm{const},\label{eq:process function}
\end{equation}
where $\gamma\equiv1+Nk_{B}/C_{g}$ is the heat capacity ratio of
the ideal gas. The above equation can be re-written in terms of $P$
and $V$ as

\begin{equation}
PV^{\xi}=\mathrm{const},\;\xi\equiv\frac{\gamma+C_{r}/C_{g}}{1+C_{r}/C_{g}}.\label{eq:PVxi}
\end{equation}
Arrive here, the polytropic process equation is derived from our simple
model, and we see that the polytropic exponent $\xi$ is determined
by the heat capacity ratio between the reservoir and gas. As an illustration,
the $P-V$ diagram of the diatomic molecule gas ($\gamma=1.4$) with
different $C_{r}/C_{g}$ is plotted in Fig. \ref{fig:-diagram-of}.
In addition, from the perspective of potential application, if the
heat capacity of the reservoir is unknown, with the polytropic exponent
obtained by measuring the $P-V$ diagram of the gas, Eq. (\ref{eq:PVxi})
indicates that the heat capacity of the reservoir can be estimated
as $C_{r}=C_{g}(\gamma-\xi)/(\xi-1)$.

We further analyze the obtained polytropic process equation in different
limits of $C_{r}$ with given $C_{g}$:

\textbf{i)} In the limit that the reservoir size tends to infinity
with $C_{r}\rightarrow\infty$, one has $\xi\rightarrow1$, and thus
$T=\mathrm{const}$, which means the process becomes an isothermal
process (red dashed curve in Fig. \ref{fig:-diagram-of}). This result
is very intuitive, because when the heat capacity of the reservoir
tends to infinity, its temperature change tends to 0 correspondingly,
that is, the reservoir becomes a heat bath with constant temperature.

\textbf{ii)} In the limit of $C_{r}\rightarrow0$, the reservoir can
be considered as disappearing, such that the gas becomes an isolated
system with $\xi\rightarrow\gamma$. In this case, Eq. (\ref{eq:process function})
recovers the adiabatic equation of ideal gas $PV^{\gamma}=\mathrm{const}$,
which is represented by the black solid curve in Fig. \ref{fig:-diagram-of}. 

It is worth mentioning that, conventional materials generally have
positive heat capacity ($C_{r}>0$), and thus $C_{r}/C_{g}\in(0,\infty)$.
In this case, the polytropic exponent $\xi\in(1,\gamma)$, which result
in Eq. (\ref{eq:PVxi}) can not describe the isochoric process ($\xi\rightarrow\infty$)
and the isobaric process ($\xi\rightarrow0$). Interestingly, it is
easy to check that the process equations of these two typical processes
can be covered by Eq. (\ref{eq:PVxi}) once the reservoir's heat capacity
can take negative values \footnote{$C_{r}=-\gamma C_{g}$ result in $\xi\rightarrow0$ while $\xi\rightarrow\infty$
is achieved with $C_{r}=-C_{g}$}. The requirement that the reservoir has negative heat capacity can
be achieved with unconventional materials \citep{schmidt2001negative,2020-finite-size}
or by applying work on the reservoir \footnote{Considering that the work is applied to the reservoir through a generalized
force $Y$ conjugated to the generalized displacement $X$ as $\delta W=YdX$,
the law of energy conservation for the reservoir in this situation
reads $dU_{r}=\delta Q_{r}+YdX$. The heat capacity of the reservoir
follows as $C_{r}'=\delta Q_{r}/dT=C_{r}-YdX/dT$, which can be negative
when $C_{r}<YdX/dT$.}, which will not be discussed in detail here.

\textit{Work and heat.--} Considering the volume of the gas with
initial temperature $T_{0}$ is tuned from $V_{0}$ to $V_{f}$ in
the discussed process, it follows from Eq. (\ref{eq:process function})
that the final temperature of the gas is

\begin{equation}
T_{f}=T_{0}\left(\frac{V_{0}}{V_{f}}\right)^{\xi-1}\equiv T_{0}\mathrm{e}^{-\lambda\kappa}
\end{equation}
with $\lambda=\ln(V_{f}/V_{0})$ and $\kappa=\xi-1=(\gamma-1)/(1+C_{r}/C_{g})$.
According to Eq. (\ref{eq:DE}), the output work of the gas is,

\begin{align}
W & =\int_{V_{0}}^{V_{f}}PdV=-\left(C_{g}+C_{r}\right)\int_{T_{0}}^{T_{f}}dT\\
 & =\left(C_{g}+C_{r}\right)T_{0}\left(1-\mathrm{e}^{-\lambda\kappa}\right).\label{eq:W}
\end{align}
In addition, it follows from Eq. (\ref{eq:dQg}) that the heat absorbed
of the gas 

\begin{equation}
Q_{g}=-C_{r}\int_{T_{0}}^{T_{f}}dT=\frac{C_{r}}{C_{g}+C_{r}}W,
\end{equation}
which is proportional to the output work. This is in consistent with
a recent study on the polytropic process \citep{christians2012approach},
where the authors defined the energy transfer ratio $\delta Q_{g}/\delta W\equiv K$
and introduced a basic assumption that $K$ is a constant.

When the heat capacity of the reservoir is much lager than that the
gas, i.e., $C_{g}\ll C_{r}$, keeping to the first order of $\kappa$
or $C_{g}/C_{r}$, the output work in Eq. (\ref{eq:W}) is approximated
as

\begin{align}
W & =\frac{Nk_{B}}{\kappa}T_{0}\left\{ 1-\left[1-\lambda\kappa+\frac{1}{2}\lambda^{2}\kappa^{2}+O\left(\kappa^{3}\right)\right]\right\} \\
 & \approx Nk_{B}T_{0}\left(\lambda-\frac{1}{2}\lambda^{2}\kappa\right)\\
 & \approx W_{\mathrm{iso}}\left(1-\frac{\gamma-1}{2C_{r}/C_{g}}\right),\label{eq:W-C}
\end{align}
where $W_{\mathrm{iso}}\equiv Nk_{B}T_{0}\ln(V_{f}/V_{0})$ is the
output work of the gas in the isothermal process with temperature
$T_{0}$. The second term in Eq. (\ref{eq:W-C}) is the correction
due to the finiteness in size of the reservoir, which will reduce
the amount of output work compared to $W_{\mathrm{iso}}$ in the case
with infinite heat reservoir. This $1/C_{r}$ correction agrees well
with some recent studies on finite-system thermodynamics and statistics
\citep{reeb2015tight,richens2018finite,timpanaro2020landauer,quantum2nd-2021ma}

\textit{Conclusion and discussions.--} The simple gas-reservoir model
proposed in this paper can deepen students' understanding of the polytropic
process and can be directly extended to the van der Waals gas system
\citep{malic1955equation}.

In common sense, the thermal reservoir is generally considered as
a thermal equilibrium system with infinite degrees of freedom, and
the temperature is its only characteristic quantity. Correction related
to the finiteness of the reservoirs in the energy conversion process
will lead students to think about the novel thermodynamic effects
off the thermodynamic limit \citep{ondrechen1981maximum,leff1987available,2020-finite-size,quantum2nd-2021ma,yuan2022optimizing}.
When the heat capacity of the thermal reservoir is temperature -dependent,
the solution of Eq. (\ref{eq:DE}) will no longer satisfy the polytropic
process equation, the $P-V$ diagram of the gas in this case is worth
further exploring. Moreover, without the quasi-static assumption,
taking into account the non-equilibrium heat transfer, this model
can also be utilized to study the irreversible thermodynamic behavior
of the gas driven in finite time \citep{CA,andresen1983finite,BroeckPRL2005,2020IEGyhma,Constraintrelationyhma}.

As a final remark, it is feasible in principle to experimentally demonstrate
our model at the undergraduate level. For example, when the ideal
gas is specific as $10$L of air ($C_{g}\approx\mathrm{9.3J/K,\gamma\approx1.4})$
around the zoom temperature, \textbf{a)} if the reservoir is 2L of
air ($C_{r}\approx\mathrm{1.86J/K})$, according to Eq. (\ref{eq:process function}),
the polytropic exponent $\xi\approx1.333$; \textbf{b)} if the reservoir
is also$10$L of air, the polytropic exponent $\xi\approx1.2$; \textbf{c)}
if the reservoir is specific as a cup of water ($200$mL, $C_{r}\approx\mathrm{840J/K}$),
the polytropic exponent $\xi\approx1.004$, which means the designed
process is very close to an isothermal process. The $P-V$ diagram
of the air in the whole process can be obtained by directly measuring
the volume and pressure of the air \citep{2020IEGyhma}. Apart from
exchanging heat with each other, isolating the gas and the reservoir
from the outside environment to avoid extra heat dissipation is the
main difficulty of this experiment. 

\textit{Acknowledgments.}-- The author would like to thank Shi-Gang
Ou, Shan-He Su, Tao Li and Hong Yuan for valuable comments on the
manuscript. This work is supported by the National Natural Science
Foundation of China (Grants No. 12088101, No. U1930402, and No. U1930403),
and the China Postdoctoral Science Foundation (Grant No. BX2021030).

\bibliographystyle{apsrev}
\bibliography{TOFS}

\begin{thebibliography}{20}
\expandafter\ifx\csname natexlab\endcsname\relax\def\natexlab#1{#1}\fi
\expandafter\ifx\csname bibnamefont\endcsname\relax
  \def\bibnamefont#1{#1}\fi
\expandafter\ifx\csname bibfnamefont\endcsname\relax
  \def\bibfnamefont#1{#1}\fi
\expandafter\ifx\csname citenamefont\endcsname\relax
  \def\citenamefont#1{#1}\fi
\expandafter\ifx\csname url\endcsname\relax
  \def\url#1{\texttt{#1}}\fi
\expandafter\ifx\csname urlprefix\endcsname\relax\def\urlprefix{URL }\fi
\providecommand{\bibinfo}[2]{#2}
\providecommand{\eprint}[2][]{\url{#2}}

\bibitem[{\citenamefont{Callen}(1960)}]{callen1960thermodynamics}
\bibinfo{author}{\bibfnamefont{H.~B.} \bibnamefont{Callen}},
  \emph{\bibinfo{title}{Thermodynamics and an Introduction to
  Thermostatistics}} (\bibinfo{publisher}{John Wiley \& Sons, New York},
  \bibinfo{year}{1960}).

\bibitem[{\citenamefont{Cengel et~al.}(2011)\citenamefont{Cengel, Boles, and
  Kano{\u{g}}lu}}]{cengel2011thermodynamics}
\bibinfo{author}{\bibfnamefont{Y.~A.} \bibnamefont{Cengel}},
  \bibinfo{author}{\bibfnamefont{M.~A.} \bibnamefont{Boles}}, \bibnamefont{and}
  \bibinfo{author}{\bibfnamefont{M.}~\bibnamefont{Kano{\u{g}}lu}},
  \emph{\bibinfo{title}{Thermodynamics: an engineering approach}},
  vol.~\bibinfo{volume}{5} (\bibinfo{publisher}{McGraw-hill New York},
  \bibinfo{year}{2011}).

\bibitem[{Note1()}]{Note1}
Note1, \bibinfo{note}{$C_{r}=-\gamma C_{g}$ result in $\xi \rightarrow 0$ while
  $\xi \rightarrow \infty $ is achieved with $C_{r}=-C_{g}$}.

\bibitem[{\citenamefont{Schmidt et~al.}(2001)\citenamefont{Schmidt, Kusche,
  Hippler, Donges, Kronm{\"u}ller, Von~Issendorff, and
  Haberland}}]{schmidt2001negative}
\bibinfo{author}{\bibfnamefont{M.}~\bibnamefont{Schmidt}},
  \bibinfo{author}{\bibfnamefont{R.}~\bibnamefont{Kusche}},
  \bibinfo{author}{\bibfnamefont{T.}~\bibnamefont{Hippler}},
  \bibinfo{author}{\bibfnamefont{J.}~\bibnamefont{Donges}},
  \bibinfo{author}{\bibfnamefont{W.}~\bibnamefont{Kronm{\"u}ller}},
  \bibinfo{author}{\bibfnamefont{B.}~\bibnamefont{Von~Issendorff}},
  \bibnamefont{and}
  \bibinfo{author}{\bibfnamefont{H.}~\bibnamefont{Haberland}},
  \bibinfo{journal}{Phys. Rev. Lett.} \textbf{\bibinfo{volume}{86}},
  \bibinfo{pages}{1191} (\bibinfo{year}{2001}).

\bibitem[{\citenamefont{Ma}(2020)}]{2020-finite-size}
\bibinfo{author}{\bibfnamefont{Y.-H.} \bibnamefont{Ma}},
  \bibinfo{journal}{Entropy} \textbf{\bibinfo{volume}{22}},
  \bibinfo{pages}{1002} (\bibinfo{year}{2020}).

\bibitem[{Note2()}]{Note2}
Note2, \bibinfo{note}{considering that the work is applied to the reservoir
  through a generalized force $Y$ conjugated to the generalized displacement
  $X$ as $\delta W=YdX$, the law of energy conservation for the reservoir in
  this situation reads $dU_{r}=\delta Q_{r}+YdX$. The heat capacity of the
  reservoir follows as $C_{r}'=\delta Q_{r}/dT=C_{r}-YdX/dT$, which can be
  negative when $C_{r}<YdX/dT$.}

\bibitem[{\citenamefont{Christians}(2012)}]{christians2012approach}
\bibinfo{author}{\bibfnamefont{J.}~\bibnamefont{Christians}},
  \bibinfo{journal}{International Journal of Mechanical Engineering Education}
  \textbf{\bibinfo{volume}{40}}, \bibinfo{pages}{53} (\bibinfo{year}{2012}).

\bibitem[{\citenamefont{Reeb and Wolf}(2015)}]{reeb2015tight}
\bibinfo{author}{\bibfnamefont{D.}~\bibnamefont{Reeb}} \bibnamefont{and}
  \bibinfo{author}{\bibfnamefont{M.~M.} \bibnamefont{Wolf}},
  \bibinfo{journal}{IEEE Transactions on Information Theory}
  \textbf{\bibinfo{volume}{61}}, \bibinfo{pages}{1458} (\bibinfo{year}{2015}).

\bibitem[{\citenamefont{Richens et~al.}(2018)\citenamefont{Richens, Alhambra,
  and Masanes}}]{richens2018finite}
\bibinfo{author}{\bibfnamefont{J.~G.} \bibnamefont{Richens}},
  \bibinfo{author}{\bibfnamefont{{\'A}.~M.} \bibnamefont{Alhambra}},
  \bibnamefont{and} \bibinfo{author}{\bibfnamefont{L.}~\bibnamefont{Masanes}},
  \bibinfo{journal}{Phys. Rev. E} \textbf{\bibinfo{volume}{97}},
  \bibinfo{pages}{062132} (\bibinfo{year}{2018}).

\bibitem[{\citenamefont{Timpanaro et~al.}(2020)\citenamefont{Timpanaro, Santos,
  and Landi}}]{timpanaro2020landauer}
\bibinfo{author}{\bibfnamefont{A.~M.} \bibnamefont{Timpanaro}},
  \bibinfo{author}{\bibfnamefont{J.~P.} \bibnamefont{Santos}},
  \bibnamefont{and} \bibinfo{author}{\bibfnamefont{G.~T.} \bibnamefont{Landi}},
  \bibinfo{journal}{Phys. Rev. Lett.} \textbf{\bibinfo{volume}{124}},
  \bibinfo{pages}{240601} (\bibinfo{year}{2020}).

\bibitem[{\citenamefont{Ma et~al.}(2021)\citenamefont{Ma, Liu, and
  Sun}}]{quantum2nd-2021ma}
\bibinfo{author}{\bibfnamefont{Y.-H.} \bibnamefont{Ma}},
  \bibinfo{author}{\bibfnamefont{C.~L.} \bibnamefont{Liu}}, \bibnamefont{and}
  \bibinfo{author}{\bibfnamefont{C.~P.} \bibnamefont{Sun}},
  \bibinfo{journal}{arXiv:2110.04550}  (\bibinfo{year}{2021}).

\bibitem[{\citenamefont{Malic}(1955)}]{malic1955equation}
\bibinfo{author}{\bibfnamefont{D.}~\bibnamefont{Malic}},
  \bibinfo{journal}{Journal of the Franklin Institute}
  \textbf{\bibinfo{volume}{259}}, \bibinfo{pages}{235} (\bibinfo{year}{1955}).

\bibitem[{\citenamefont{Ondrechen et~al.}(1981)\citenamefont{Ondrechen,
  Andresen, Mozurkewich, and Berry}}]{ondrechen1981maximum}
\bibinfo{author}{\bibfnamefont{M.~J.} \bibnamefont{Ondrechen}},
  \bibinfo{author}{\bibfnamefont{B.}~\bibnamefont{Andresen}},
  \bibinfo{author}{\bibfnamefont{M.}~\bibnamefont{Mozurkewich}},
  \bibnamefont{and} \bibinfo{author}{\bibfnamefont{R.~S.} \bibnamefont{Berry}},
  \bibinfo{journal}{Am. J. Phys.} \textbf{\bibinfo{volume}{49}},
  \bibinfo{pages}{681} (\bibinfo{year}{1981}).

\bibitem[{\citenamefont{Leff}(1987)}]{leff1987available}
\bibinfo{author}{\bibfnamefont{H.~S.} \bibnamefont{Leff}},
  \bibinfo{journal}{Am. J. Phys.} \textbf{\bibinfo{volume}{55}},
  \bibinfo{pages}{701} (\bibinfo{year}{1987}).

\bibitem[{\citenamefont{Yuan et~al.}(2022)\citenamefont{Yuan, Ma, and
  Sun}}]{yuan2022optimizing}
\bibinfo{author}{\bibfnamefont{H.}~\bibnamefont{Yuan}},
  \bibinfo{author}{\bibfnamefont{Y.-H.} \bibnamefont{Ma}}, \bibnamefont{and}
  \bibinfo{author}{\bibfnamefont{C.}~\bibnamefont{Sun}},
  \bibinfo{journal}{Phys. Rev. E} \textbf{\bibinfo{volume}{105}},
  \bibinfo{pages}{L022101} (\bibinfo{year}{2022}).

\bibitem[{\citenamefont{Curzon and Ahlborn}(1975)}]{CA}
\bibinfo{author}{\bibfnamefont{F.~L.} \bibnamefont{Curzon}} \bibnamefont{and}
  \bibinfo{author}{\bibfnamefont{B.}~\bibnamefont{Ahlborn}},
  \bibinfo{journal}{Am. J. Phys.} \textbf{\bibinfo{volume}{43}},
  \bibinfo{pages}{22} (\bibinfo{year}{1975}).

\bibitem[{\citenamefont{Andresen}(1983)}]{andresen1983finite}
\bibinfo{author}{\bibfnamefont{B.}~\bibnamefont{Andresen}},
  \emph{\bibinfo{title}{Finite-time thermodynamics}}
  (\bibinfo{publisher}{University of Copenhagen Copenhagen},
  \bibinfo{year}{1983}).

\bibitem[{\citenamefont{den Broeck}(2005)}]{BroeckPRL2005}
\bibinfo{author}{\bibfnamefont{C.~V.} \bibnamefont{den Broeck}},
  \bibinfo{journal}{Phys. Rev. Lett.} \textbf{\bibinfo{volume}{95}},
  \bibinfo{pages}{190602} (\bibinfo{year}{2005}).

\bibitem[{\citenamefont{Ma et~al.}(2020)\citenamefont{Ma, Zhai, Chen, Dong, and
  Sun}}]{2020IEGyhma}
\bibinfo{author}{\bibfnamefont{Y.-H.} \bibnamefont{Ma}},
  \bibinfo{author}{\bibfnamefont{R.-X.} \bibnamefont{Zhai}},
  \bibinfo{author}{\bibfnamefont{J.}~\bibnamefont{Chen}},
  \bibinfo{author}{\bibfnamefont{H.}~\bibnamefont{Dong}}, \bibnamefont{and}
  \bibinfo{author}{\bibfnamefont{C.~P.} \bibnamefont{Sun}},
  \bibinfo{journal}{Phys. Rev. Lett.} \textbf{\bibinfo{volume}{125}},
  \bibinfo{pages}{210601} (\bibinfo{year}{2020}).

\bibitem[{\citenamefont{Ma et~al.}(2018)\citenamefont{Ma, Xu, Dong, and
  Sun}}]{Constraintrelationyhma}
\bibinfo{author}{\bibfnamefont{Y.-H.} \bibnamefont{Ma}},
  \bibinfo{author}{\bibfnamefont{D.}~\bibnamefont{Xu}},
  \bibinfo{author}{\bibfnamefont{H.}~\bibnamefont{Dong}}, \bibnamefont{and}
  \bibinfo{author}{\bibfnamefont{C.-P.} \bibnamefont{Sun}},
  \bibinfo{journal}{Phys. Rev. E} \textbf{\bibinfo{volume}{98}},
  \bibinfo{pages}{042112} (\bibinfo{year}{2018}).

\end{thebibliography}

\end{document}